\documentclass[a4paper]{llncs}
\usepackage{amsmath,amssymb}
\usepackage{graphics}
\usepackage{graphicx}
\usepackage{hyperref}
\usepackage{subfigure}
\usepackage[ruled,vlined]{algorithm2e}

\usepackage{url}
\urldef{\mailsa}\path|{alfred.hofmann, ursula.barth, ingrid.haas, frank.holzwarth,|
\urldef{\mailsb}\path|anna.kramer, leonie.kunz, christine.reiss, nicole.sator,|
\urldef{\mailsc}\path|erika.siebert-cole, peter.strasser, lncs}@springer.com|

\begin{document}
\mainmatter
\title{Automated Change Detection and Reactive Clustering in Multivariate Streaming Data}
\author{ Dang-Hoan Tran}
\institute{Vietnam Maritime University, Vietnam\\ \email{hoantd@gmail.com} }

\maketitle
\begin{abstract}
 Many automated systems need the capability of automatic change detection without the given detection threshold. This paper presents an automated change detection algorithm in streaming multivariate data.

Two overlapping windows are used to quantify the changes. While a window is used as the reference window from which the clustering is created, the other called the current window captures the newly incoming data points. A newly incoming data point can be considered a change point if it is not a member of any cluster. As our clustering-based change detector does not require detection threshold, it is an automated detector. Based on this change detector, we propose a reactive clustering algorithm for streaming data.

Our empirical results show that, our clustering-based change detector works well with multivariate streaming data. The detection accuracy depends on the number of clusters in the reference window, the window width.
\end{abstract}

\section{Introduction}
\label{sec:introduction}
Change detection in data streams is an important and active area of research in both data stream processing and data stream mining. Detecting and monitoring changes in streaming data can detect and react to the interesting events. We study the problem of change detection in streaming multivariate data, and building and maintenance of clustering built from streaming data by using the reactive approach.

Change detection in streaming multivariate data is important because of the following reasons.
First, event detection is often based on change detection in multivariate streaming data. For example, early detection of fire is detected  based on increase in temperature and light intensity, and decrease in humidity.
Second, as the environment changes, change detection methods need the capability of automatically detecting changes.
Third, as nearby sensors may have high correlation in readings, the changes in incoming data stream lead to the clustering evolution.

\begin{figure}
    \centering
         \includegraphics[width=0.40\textwidth]{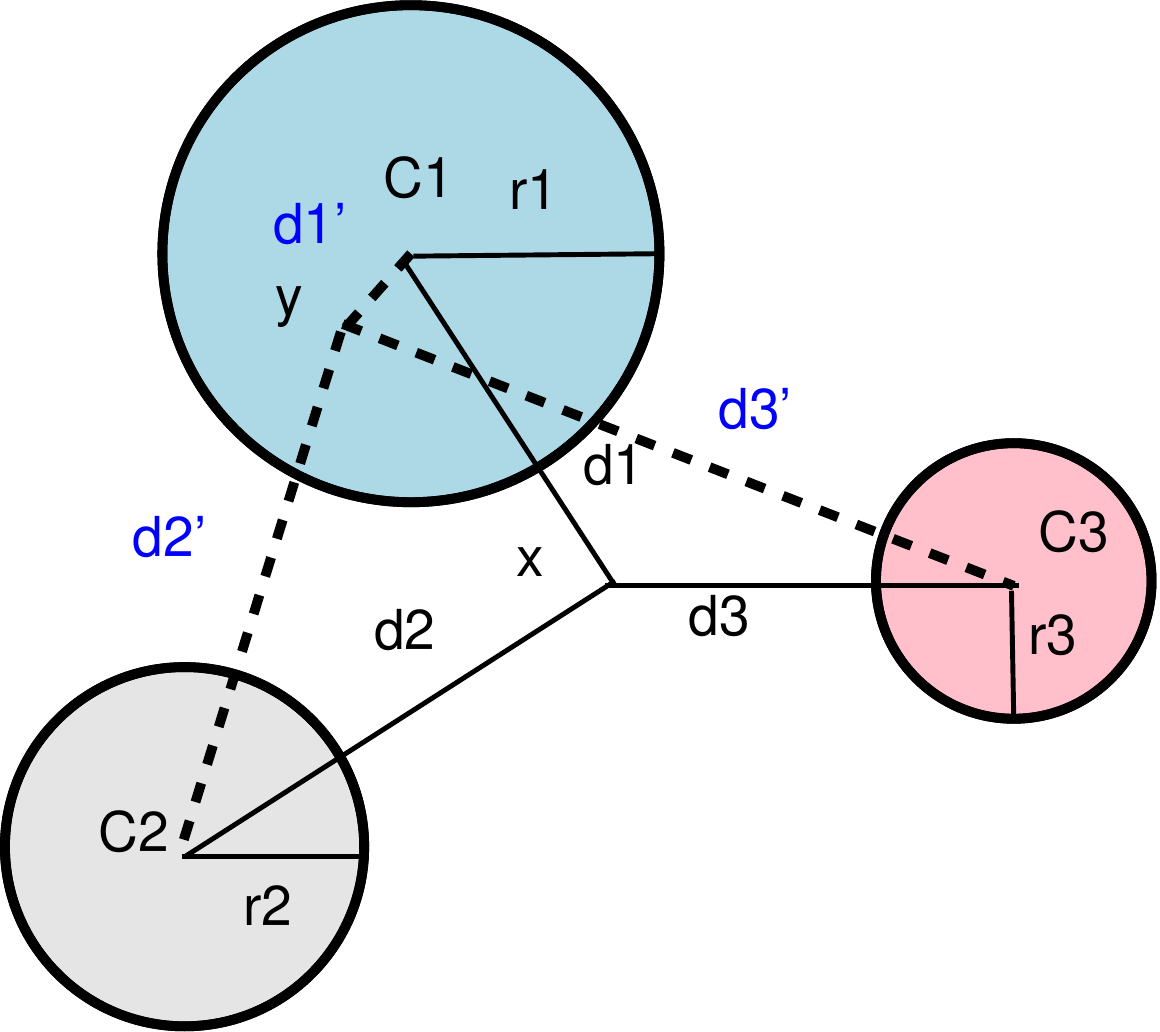}
    \caption{Change detection by clustering}
    \label{fig:automaticthreshold}
\end{figure}

Our change detection method uses the model fitting approach in which a change can be considered to occur when a new data item or block of data items  do not fit the existing clustering. Figure \ref{fig:automaticthreshold} illustrates how the clustering-based method for detecting changes works. There is a clustering of three clusters $C1, C2, C3$ in the reference window. Let $x$ or $y$ be newly incoming data items. As all the distances from data point $x$ to three centers of clusters $C1, C2, C3$ are greater than the radiuses of three corresponding clusters, data point $x$ is a change point while data point $y$ is non-change point, because it is a member of cluster $C1$. 

Building models from the continuous data streams aims to capture patterns and time-evolving trends in these streams. There are three approaches to building and monitoring models extracted from data streams: periodic, incremental, and reactive approaches \cite{bifet2010adaptive}. In the periodic approach, the model is rebuilt time to time. The periodic approach may incur high cost in terms of resources and model accuracy. In particular, this approach may not well be suited for streaming evolving data. In the incremental approach, the model is updated whenever the data changes. The benefits of incremental approaches are accurate and optimal. The reactive approach  monitors the change, and rebuilds the model only when it no longer suits the data. The best approach to building and maintaining model is the incremental approach. However, this incremental approach depends every specific problem. The reactive approach is widely used for many computation tasks because of its simplicity and efficiency \cite{bhaduri2008efficient}.

The contributions of this paper are as follows
\begin{itemize}
\item We present an automated change detection method in multivariate streaming data by using the geometric and clustering approach.
\item  Based on the above method for change detection, we present method for building and maintaining of clustering by using the reactive approach in which this clustering-based change detection method is used to determine when to rebuild clustering.
\end{itemize}

We formulate criteria for detecting changes and  describe the algorithms in in Section \ref{lbl:modelcdformulation}.  The experimental results, as well as evaluations of our framework are presented in  Section \ref{sec:eval}.  Related work is given in Section \ref{sec:relatedwork}. Finally, we conclude in Section \ref{sec:Summary}.

\section{Automated Detection and Reactive Clustering}\label{lbl:modelcdformulation}
This section formulates the criteria for detecting changes in multivariate streaming data by using clustering and geometric approach.
A data stream is an infinite sequence of elements
\begin{equation}
S=\left\{ \left(X_{1},T_{1}\right),..,\left(X_{i},T_{i}\right),...\right\}
\end{equation}
Each element is a pair $\left(X_{i},\, T_{i}\right)$ where $X_{i}$ is a $d$-dimensional vector $X_{i}=\left(x_{1},\, x_{2},...,\,x_{d}\right)$ arriving at the time stamp $T_{i}$.
As change detection is process of identifying differences in the state of an object or phenomenon by observing it at different times and/or different locations in space, change detection can reduce to the problem of hypothesis testing below
\begin{equation}
\begin{cases}
\mathcal{H}_{0} & \neg change\\
\mathcal{H}_{1} & change
\end{cases}
\end{equation}
where the hypothesis is a logical expression $change$ which indicates whether a change occurs.
Result of a continuous change detector in streaming data  can be considered a sequence of  results of  one-time change detector.

We now present an automated method for detecting changes in multivariate streaming data by using clustering and geometric approach. It is an automated change detection method because it does not require the given threshold. As we use K-means to produce clustering, and the Euclidean distance as similarity measure \cite{jain2010data}, the clusters are spherically-shaped clusters.

Let $X$ denote a recently arriving data item. A data point can be considered a change point if it is not a member of any cluster. In other words, data point $X$ is a change point if the following logical expression is true
\begin{equation}\label{eq:changecondition1}
change=\underset{i=1}{\overset{K}{\wedge}}\left[d\left(X,center\left(C_{i}\right)\right)>radius\left(C_{i}\right)\right]
\end{equation}
where $d\left(X,center\left(C_{i}\right)\right)$ is the distance between a newly incoming data point $X$ and the center of cluster $C_{i}$, and $radius\left(C_{i}\right)$ is the radius of the cluster $C_{i}$, for $i \in {1,..,K}$, where $K$ is the number of clusters. The centroid of a cluster is computed by formula \ref{eq:centroid}. The radius of each cluster is computed by formula \ref{eq:radius}. The distance between a new data point and the centroid of the cluster $i$ is computed by formular \ref{eq:dist}.

Expression \ref{eq:changecondition1} checks whether a data point is a change point. However, it can be only used for one incoming data point. We now design a logical expression for checking whether the changes occur when sliding step is a block of $b$ data points $X_1,X_2,..,X_b$. A block of data points is considered change if at least one change point belongs to this block. Therefore, the logical expression for checking whether a block of data points change in comparison with the reference window is as follows.
\begin{equation}\label{eq:changecondition2}
change\left(X_{1},..,X_{b}\right)=change\left(X_{1}\right)\vee ..\vee change\left(X_{b}\right)
\end{equation}
where $change\left(X_{i}\right)$ is determined by Expression \ref{eq:changecondition1}, and $1\leq i\leq b$.



The first truly scalable algorithm for clustering data stream called BIRCH \cite{zhang1996birch} constructs a clustering structure in a single scan over the data with limited memory. BIRCH can work with limited resources. The characteristics of BIRCH includes incremental clustering, compact representation of clusters, and the ability to process data in a single pass. With the above characteristics, BIRCH is well suited for clustering large database as well as the evolving data streams. The underlying concept behind BIRCH called the cluster feature vector is defined as follows
\begin{definition}\label{defn: cf}
Clustering Feature \cite{zhang1996birch} Given d-dimensional data
points in a cluster: $\left\{ \vec{X}\right\} $ where $i=1,2,..,N$,
the Clustering Feature (CF) vector of the cluster is a triple: $CF=\left(N,\, LS,\, SS\right)$
where $N$ is the number of data points in the cluster, $LS=\underset{i=0}{\overset{N-1}{\sum}}X_{i}$
is the linear sum of the data points in the cluster, and $SS=\underset{i=0}{\overset{N-1}{\sum}}X_{i}^{2}$
is the squared sum of the N data points. The cluster created by merging
two above disjoint clusters $CF_{1}$ and, $CF_{2}$ has the cluster feature is defined as
follows
\begin{equation}
CF=\left(N_{1}+N_{2},\, LS_{1}+LS_{2},\, SS_{1}+SS_{2}\right)
\end{equation}
\end{definition}
From cluster feature vector, the centroid and the radius of the cluster are computed as follows
\begin{equation}\label{eq:centroid}
{X_0} = \frac{{\sum\limits_{i = 1}^N {{X_i}} }}{N}
\end{equation}
and
\begin{equation}\label{eq:radius}
R = \sqrt {\frac{{\sum\limits_{i = 1}^N {{{\left( {{X_i} - {X_0}} \right)}^2}} }}{N}}
\end{equation}
The distance from a new data point $X$ to the centroid of the cluster $i$ ${{\rm{X}}_{{{\rm{0}}_{\rm{i}}}}}$ is computed by using the Euclidean distance
\begin{equation}\label{eq:dist}
d\left( {X,{X_{{0_i}}}} \right) = \sqrt {{{\left( {{X_{{0_i}}} - X} \right)}^2}}
\end{equation}
The advantages of this CF summary are that it does not require to store all the data points in the cluster, and it provides sufficient information for computing all the measurements necessary for making clustering decisions \cite{zhang1996birch}.

Aggarwal et al. \cite{aggarwal2003framework} extended the concept cluster feature vector for streaming context by adding the temporal components.
\begin{definition}\label{defn: tempcf} Micro-cluster \cite{aggarwal2003framework}. A micro-cluster for a set of $d-$dimensional points $X_{i_{1}},...,X_{i_{N}}$
with time stamps $T_{i_{1}},...,T_{i_{n}}$ is the $\left(2d+3\right)$-tuple
\\$\left(\overline{CF2^{x}},\overline{CF1^{x}},CF2^{t},CF1^{t},N\right)$,
wherein $\overline{CF2^{x}}$ and $\overline{CF1^{x}}$ each corresponds to a vector of $d$ entries. The definition of each of these entries is as follows
\begin{itemize}
\item For each dimension, the sum of the squares of the data values is maintained in $\overline{CF2^{x}}$. Thus, $\overline{CF2^{x}}$ contains $d$ values. The $p-th$ entry of $\overline{CF2^{x}}$  is equal to $\underset{j=1}{\overset{N}{\sum}}\left(X_{i_{j}}^{p}\right)^{2}$.
\item For each dimension, the sum of the data values is maintained in $\overline{CF1^{x}}$ . Thus, $\overline{CF1^{x}}$  contains $d$ values. The $p-th$ entry of $\overline{CF1^{x}}$  is equal to $\underset{j=1}{\overset{N}{\sum}}\left(X_{i_{j}}^{p}\right)$.
\item The sum of the squares of the time stamps $T_{i_{1}},...,T_{i_{n}}$ is maintained in $CF2^{t}$.
\item The sum of the time stamps $T_{i_{1}},...,T_{i_{N}}$ is maintained in $CF1^{t}$.
\item The number of data points is maintained in $N$.
\end{itemize}
\end{definition}

All the data points are encoded as micro-clusters. The center and the radius of each cluster can be computed from its micro-cluster as shown in Formulas \ref{eq:centroid}, and \ref{eq:radius}. As the value of the data decreases over time, instead of storing for later analysis, data is immediately analyzed as it is produced by using tuple-based sliding windows.  Our clustering-based change detector is based on the overlapping windows model. In particular, there are two tuple-based windows: the reference window $w1$ and the current window $w2$. The current window is used to capture new items. Algorithm \ref{alg:clusteringcd2} describes an clustering-based change detection method. It works as follows.
\begin{itemize}
  \item \emph{Initialization:} Read the first $N$ data items from the incoming stream into the reference window $w_{1}$.  The current window is the content of the reference window that slides one step to capture new data item. Windows $w1$ and $w2$ are overlapping by $N-1$ data points. The next step is building a clustering in the reference window $w1$.
  \item \emph{Continuous monitoring:} Test criteria for change as in Expression \ref{eq:changecondition1} or \ref{eq:changecondition2}. The boolean function $change(C_1,blk)$ determined by Expression \ref{eq:changecondition2} checks whether at least a change point exists in the block of data points $blk$. If a change is detected, the detector makes an alarm, and the current window becomes the reference window, construct the new clustering for the new reference window. The current window $w2$ always slides one step forward whether the recently arriving data item changes or not. 
\end{itemize}

\begin{algorithm}
\SetKwData{Left}{left}\SetKwData{This}{this}\SetKwData{Up}{up}
\SetKwInOut{Input}{input}\SetKwInOut{Output}{output}
\Input{A multivariate data stream $S$, $N$ is the width of tuple-based window, $K$ is the number of clusters}
\Output{The messages reporting the changes occurred, and stream of clusterings}
\BlankLine
\emph{1. Initialization:}
\Begin{
$t \longleftarrow 0$\;
$w1 \longleftarrow$  first $N$ points from
time t\tcc*{Each data point is encoded as a micro-cluster}

$\mathcal{C}_{1} \longleftarrow Kmeans(w1,K)$\;
Processing clustering C1\;
$w2 \longleftarrow slide(w1,1)$\;
$blk \longleftarrow newItemBlock(w2)$\;

}
\emph{2. Continuous monitoring:}
\While{not at the end of the stream}{
\If{$change(C_1,blk)$}{
$t \longleftarrow$ current time\;
Report change occurred\;
$w1 \longleftarrow  w2$\;
$\mathcal{C}_{1} \longleftarrow Kmeans(w1,K)$\;
Processing clustering C1\;
}{
$w2 \longleftarrow slide(w2,1)$\;
$blk \longleftarrow newItemBlock(w2)$\;
}
}
\caption{Clustering-based algorithm for detecting changes by using overlapping windows model and reactive clustering}\label{alg:clusteringcd2}
\end{algorithm}

Algorithm \ref{alg:clusteringcd2} also describes how to maintain a streaming clustering structure undergoing insertion, and deletion of items when the sliding window moves on the data stream. If a change  occurs, a new clustering is rebuilt. This clustering is returned as the result of some clustering query, or is stored in the history of clustering in streaming data warehouse.

\section{Evaluation}\label{sec:eval}
The clustering-based change detector was written in Java. An algorithm for detecting changes in streaming data should be evaluated in three aspects  scalability, accuracy, and monitoring capability. The detection accuracy of a change detection method depends on the window width and the number of clusters. The experiments on change detector using clustering were divided into two groups in terms of detection accuracy, running time, and  memory consumption. The first group of experiments evaluated the effectiveness of window width on the clustering-based change detector. The next group of experiments studied the effectiveness of the number of clusters on the performance of the clustering-based change detector. We analyzed the effect of the window width, number of clusters on the number of change points detected by the clustering-based detectors for change.

One of the challenges in assessing a change detection algorithm is the lack of ground truth data. In particular, the evaluation of a method for detecting changes in multivariate streaming data is more challenging. Therefore, the synthetic data was used for evaluating the performance of the change detection algorithms. The synthetic data set to evaluate the accuracy of our clustering-based change detection algorithm is the streaming data set HyperP \cite{zhu2010}. HyperP stream is a synthetic data stream of gradually evolving (drifting) concepts. The Hyper Plane stream consists of 100000 instances, and each instance consists of 10 attributes. These instances may fall into one of 5 classes (clusters).
To see the advantages of the continuous detector of change over the restarting detector of change, we ran both continuous detector and restarting detector. Restarting detection of changes means that a detector will be reset if a change is detected while non-restarting detector continuously detects the changes. Non-restarting detection of change is also called continuous change detection.

We ran the detectors by clustering for two cases. First, to see the effect of the window width on detection performance of the detectors, we fixed the number of clusters while changing the window width. Second, to see the effect of the number of clusters on the detection performance of the detectors, we fixed the window width while the number of clusters changes.

\subsection{Effectiveness of Window Width}
To study the effectiveness of window width on the performance of clustering-based change detector, the number of clusters was fixed to 5, and the window width was varied in the range {200,400,600,800,1000,1200,1400,1600,1800,200}.
\begin{figure}
   \centering
   \subfigure[Effectiveness of window width on running time of clustering-based method for detecting changes]{\includegraphics[width = 2.3 in]{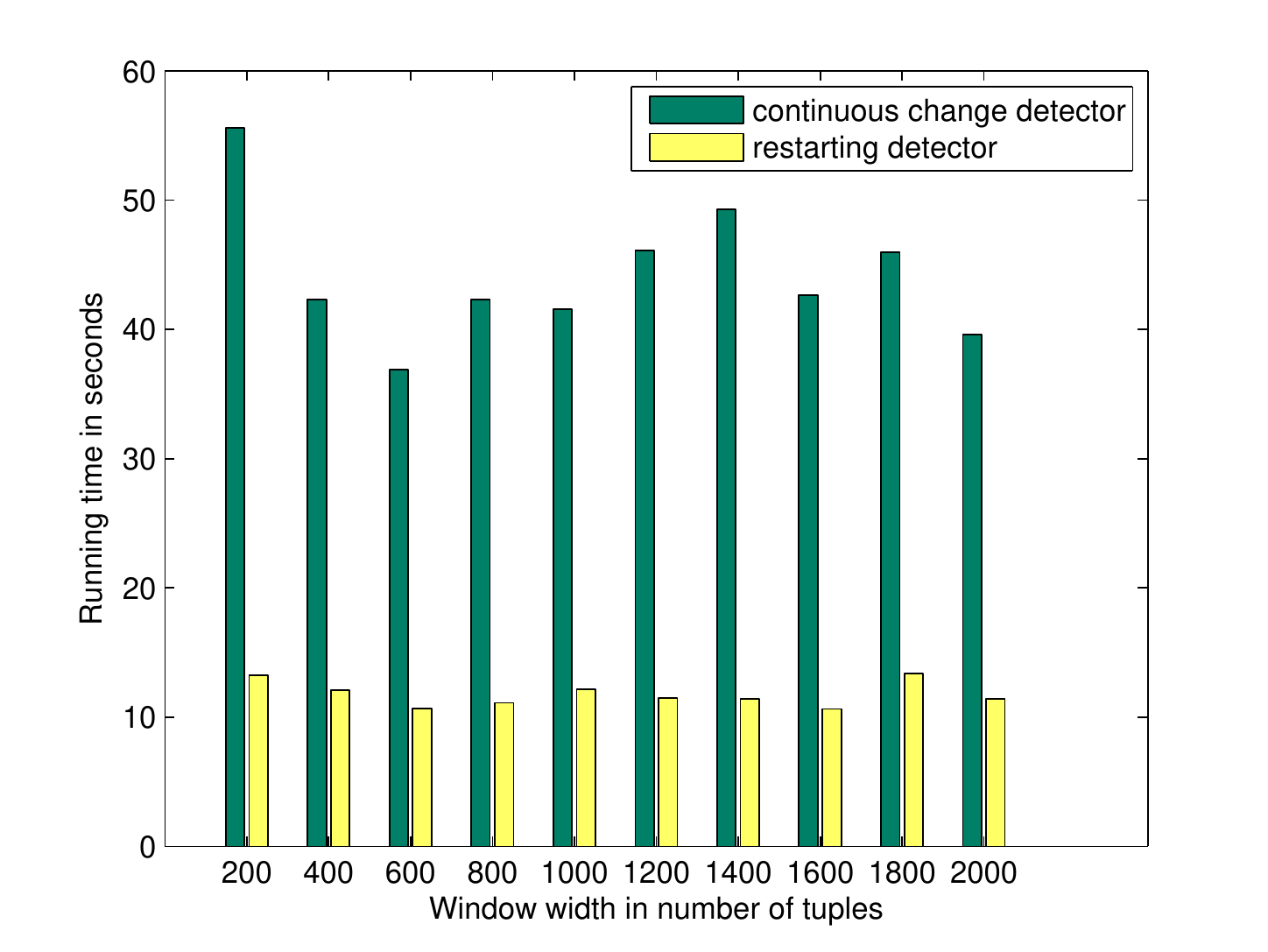}\label{fig:wineffectclusteringcd}}
   \subfigure[Effectiveness of the window width on the number of change points detected by detectors]{\includegraphics[width = 2.3 in]{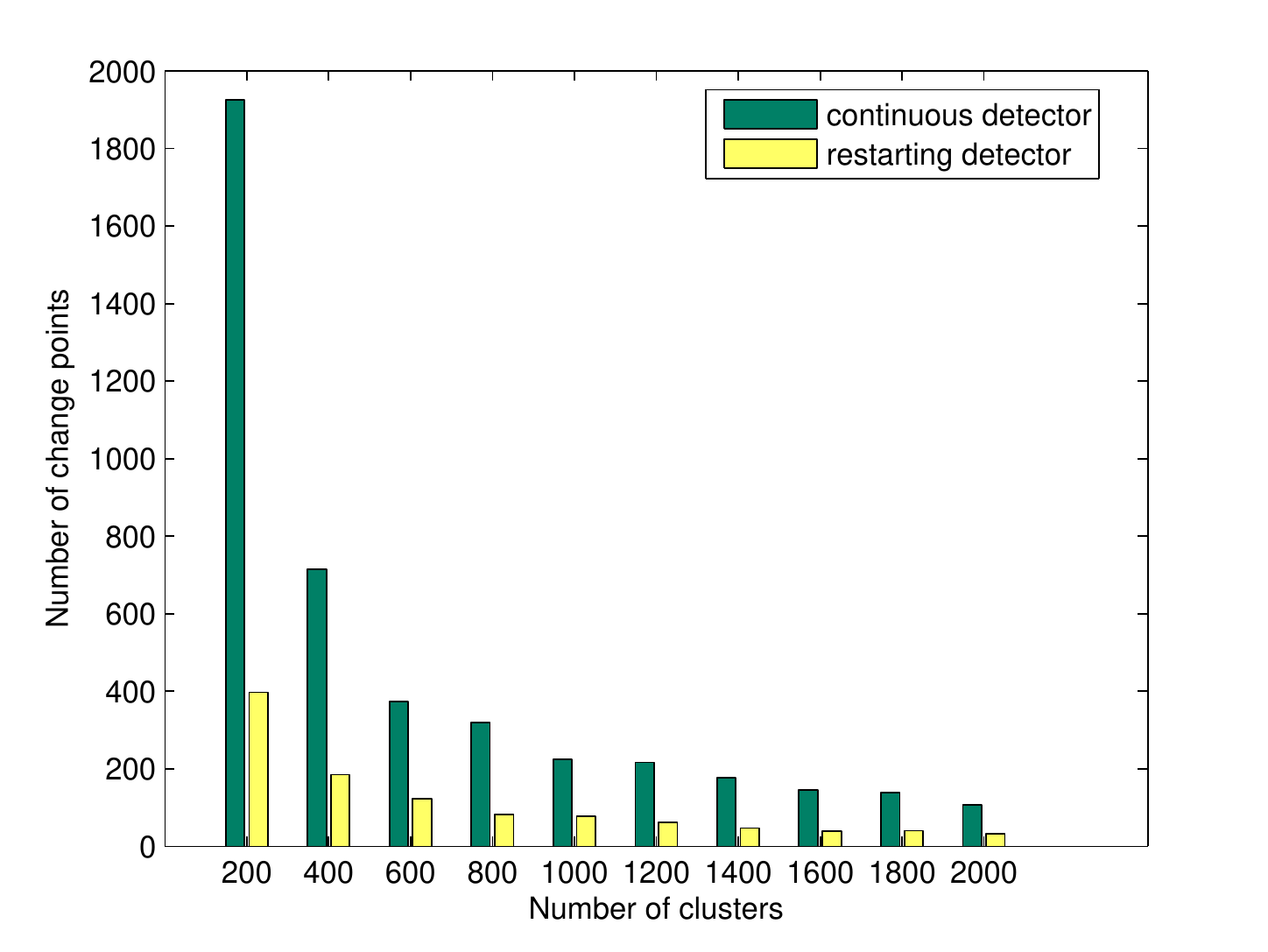}\label{fig:cpwin}}
   \caption{\label{fig:continuousdd}Effectiveness of window width on change detectors}
\end{figure}

Figure \ref{fig:wineffectclusteringcd} shows that the restarting detector runs faster than the continuous detector because the restarting detector was reset every time a change is detected. The speed of restarting detector comes at the price of its detection accuracy. As Figure \ref{fig:cpwin} shows, the number of detected change points reduced when the window width increased. The continuous detector detected much more change points than the restarting detector. Because the restarting detector was reset every time a change was detected, many points in the reference window in the resetting phase were ignored by detectors.


\subsection{Effectiveness of Cluster Number}
To assess the effectiveness of cluster number on this clustering-based change detection, the window width was fixed to 1000 tuples, and the number of clusters was varied in the range $\left(2,3,4,5,6,7,8,9,10\right)$.
\begin{figure}
   \centering
   \subfigure[Effectiveness of the cluster number on running time of the clustering-based change detection method]{\includegraphics[width = 2.3 in]{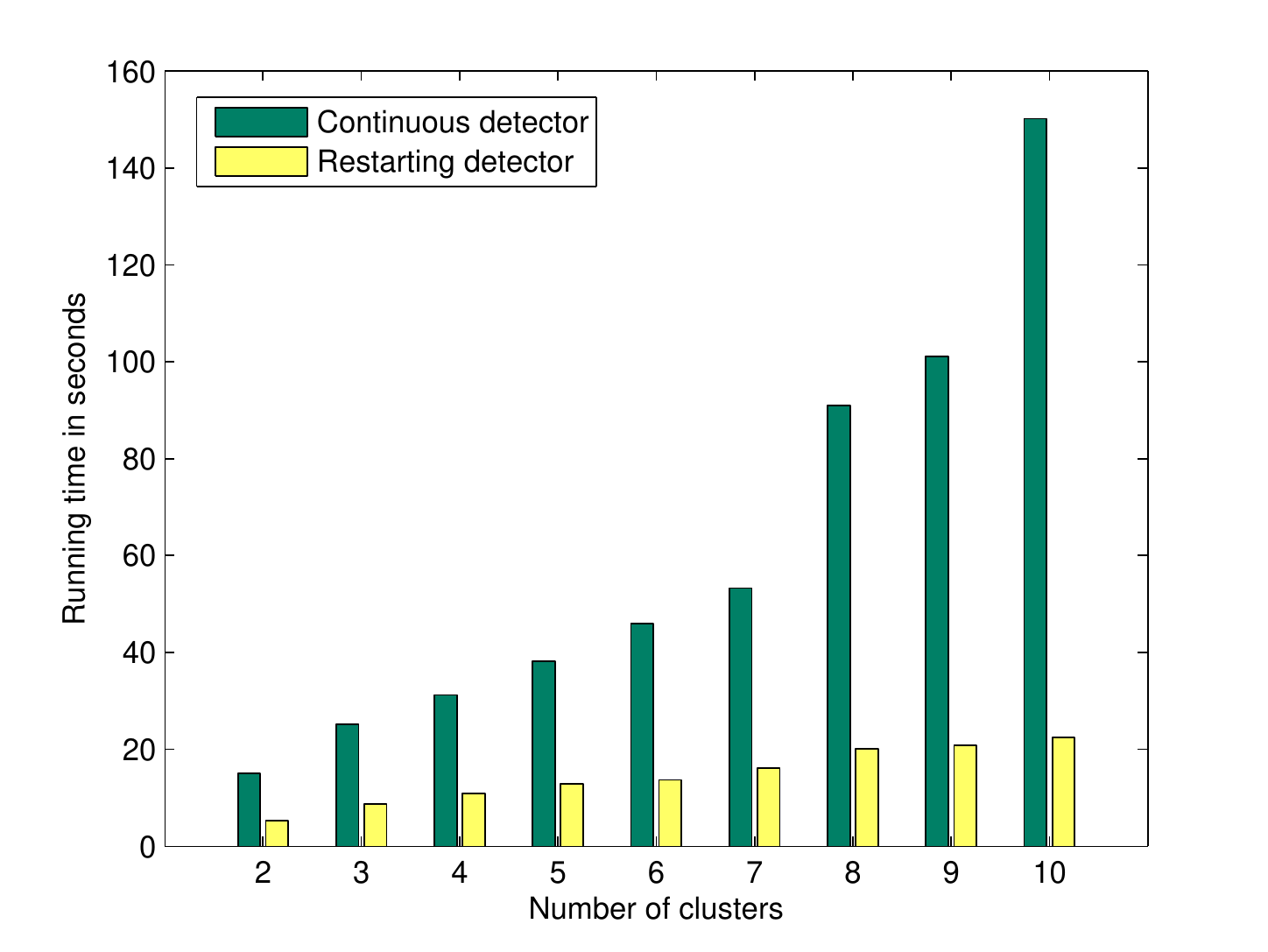}\label{fig:timecluster}}
   \subfigure[Effectiveness of number of clusters on the change points detected by the clustering-based change detector]{\includegraphics[width = 2.3 in]{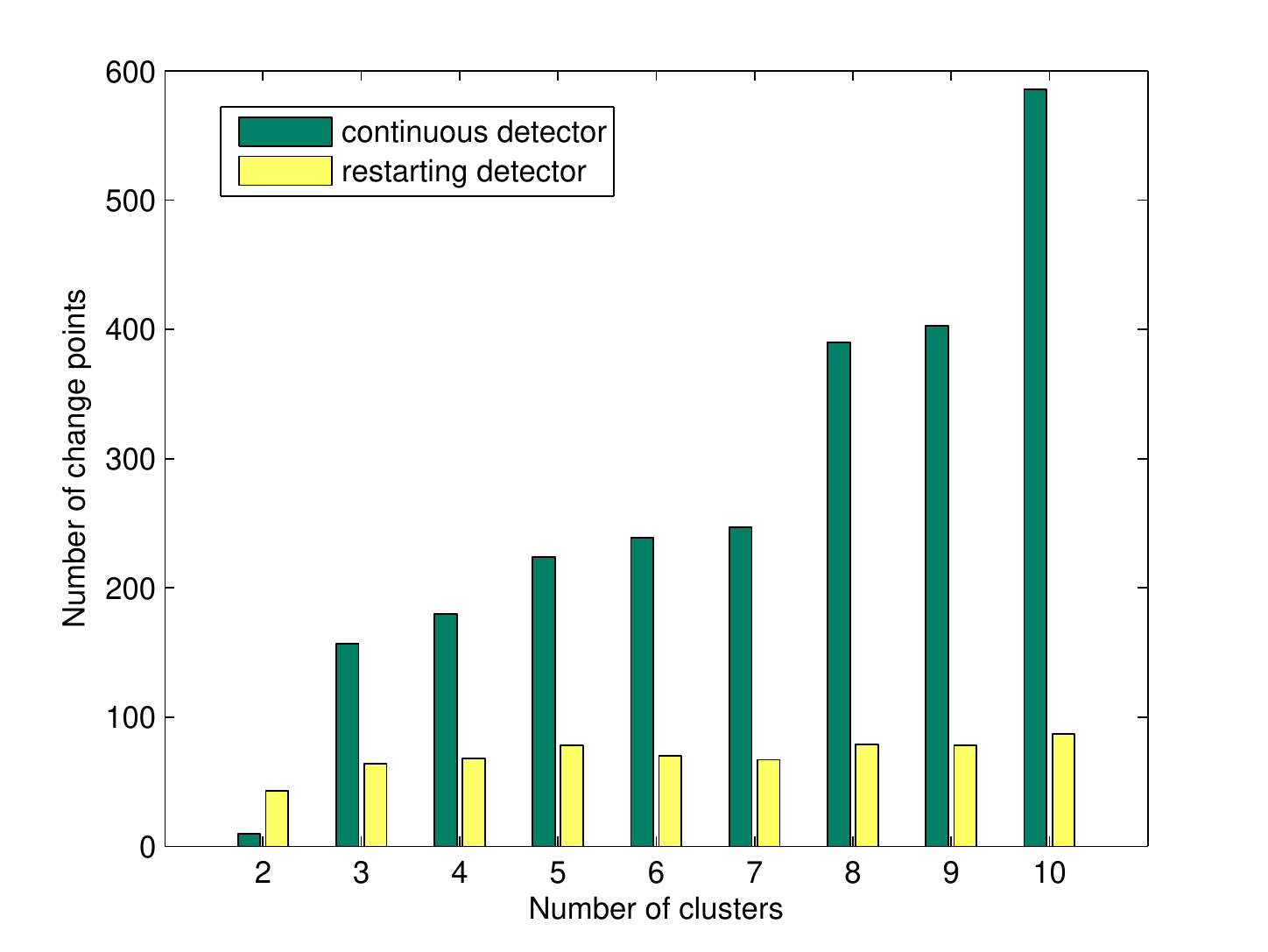}\label{fig:cpcluster}}
   \caption{\label{fig:continuousdd}Effectiveness of number of clusters on change detectors}
\end{figure}


The results of this experiment (Figure \ref{fig:timecluster}) shows that, the running time of a clustering-based change detector increases with the increasing number of clusters. The time needed to determine whether a newly incoming data point is a change point increases, as the number of comparisons between the distance from the newly incoming point to the centers of clusters and the corresponding radiuses of clusters increases with the number of clusters.

Figure \ref{fig:cpcluster} shows that, the number of change points increases when the number of clusters increases.

If the changes frequently occur, we need to run the clustering algorithm in order to create the clustering from the reference window many times.  Because the clustering process consumes a lot of time. Therefore running time of a clustering-based change detector increases with the increasing number of change points.
As such, the efficiency of the clustering-based change detection method heavily depends the selection of the number of clusters.

This could be unsuitable for streaming environment with high data speed. An alternative solution to K-means is K-means++ proposed by Arthur and  Vassilvitski \cite{arthur2007k} with the computational complexity $O\left(\log K\right)$, where $K$ is the number of centroids.

\subsection{Evaluation on Clusterings using Reactive Approach}
This group of experiments shows the effectiveness of building and maintenance of clusterings using the reactive approach on the detection performance of clustering-based change detection algorithm. The experiments were executed on the synthetic data streams HyperP. The window size was fixed to $500$, and the number of clusters was varied from 4 to 7. Table \ref{tbl:effectreactive} shows the results of interest include the number of detected change points, the number of clusterings that are generated, and the running time. The running time increases with the number of clusters because the larger number of clusters is, the more number of comparisons must be done. The number of clusterings depends on the number of clusters.
\begin{table}
\centering
\begin{tabular}{|c|c|c|c|}
\hline
Cluster number & changes & clusterings & Running time\tabularnewline
\hline
\hline
4 & 897 & 898 & 35567 ms\tabularnewline
\hline
5 & 1304 & 1305 & 54913 ms\tabularnewline
\hline
6 & 1269 & 1270 & 59422 ms\tabularnewline
\hline
7 & 1336 & 1337 & 65998 ms\tabularnewline
\hline
\end{tabular}

\caption{Effect of cluster number on the number of detected change points and
the number of clusterings}\label{tbl:effectreactive}
\end{table}
The results we want to determine include: running times of two methods, the number of clusterings that are generated. Results of method for building and maintaining clustering over sliding window using the reactive approach are as follows:  number of clusterings that are generated  is 1305;running time is 57330 ms ;used memory is bytes is 10 MB. Table \ref{tbl:effectreactive} shows that the number of clusterings generated by the reactive approach is equal to the number of change points plus one. In addition to the initial clustering, if a change occurs, a new clustering is generated. In general, the number of change points increases with the number of clusters.
\section{Related Work}\label{sec:relatedwork}
The clustering-based change detection method proposed here is related to the work on automatic change detection and change detection in multivariate streaming data, and clustering-based change detection, and the reactive work on building and maintaining of model.
As automated systems require the capability of real-time processing, and adapting to the changing environments, automated change detection plays an important role in many automated systems. One of the models of real-time processing is data stream processing.  For example, sensor networks need the automated change detection methods in which detection threshold  must adapt to the changes of the environment. Automated change detection method is also important in many mobile robotic applications \cite{neuman2011segmentation}. For example, Neuman et al. \cite{neuman2011segmentation} have proposed an online change detection method for mobile robots based on the segmentation approach. Recently Hirte et al. \cite{hirte2012data3} have developed Data3, a Kinect interface for human motion detection. In fact, Data3 is a kind of system capable of detecting the changes in spatial-temporal streaming data.

Automated systems should be capable of automatically detecting the changes without the given detection threshold. Some change detection methods can automatically tune the detection thresholds so that the rate of false alarms is not greater than a given rate of false alarms. Gustafson and Palmquist deal with the problem of automated tuning of change detectors with given false alarm rate \cite{gustafsson1997change}. Their approach computes the detection threshold by estimating a parametric distribution. The advantage of this method is that they can predict detection threshold with no or few false alarms from the used data. However, it is parametric method.

The automatic selection of threshold is of special importance.
Alippi et al. \cite{alippi2012hmm} have presented an automated change detection method for streaming data based on Hidden Markov Models. This HMM-based method is an automated change detection method by thresholding.  Their algorithm consists of the following steps: model the relationships among data streams a sequence of time invariant linear dynamic system; model the evolution of the estimated parameters of the model by Hidden Markov Model; evaluate the likelihood of new parameters; detect change based on a given threshold. If the likelihood is less than a given threshold, a change is detected. 

Most methods for detecting changes in streaming multivariate data are based on the multivariate tests. As the problem of testing statistical hypotheses in high dimensional data is particularly challenging and the change detection in streaming data requires to respond to the changes in nearly real-time, change detection in streaming multivariate data is challenging.

There are two approaches to the problem of change detection in streaming multivariate data. The first approach is based on the transformation that converts a multivariate data stream into a univariate data stream. Change detection is then performed on the univariate data stream.
Dasu et al. \cite{dasu2009change} have used this approach to design a change detection method for streaming multivariate data. In particular, a multivariate data stream is converted into a univariate data stream. The change detection task is performed by using Kolmogorov-Smirnov test. Similarly, Kim et al. \cite{kim2009using} have recently proposed the concept called the detection stream. A detection stream is a univariate stream that is generated by mapping a multivariate stream into stream of dissimilarity measures quantifying the difference between two windows. 

The second approach is developing new methods for detecting the changes in multivariate streaming data. Kuncheva has recently presented a general method for detecting changes in multivariate streaming data by using likelihood ratio-based test \cite{kuncheva2011change}. Closely related to our work is the segment-based method for detecting in multivariate data stream proposed by Chen et al. \cite{chen2009segment}. Gretton et al. \cite{gretton2012kernel} present a kernel two-sample test that can check whether two multivariate samples coming from the same distribution. Furthermore, a kernel two-sample test with the linear computational complexity suitable for streaming environment is proposed.

In contrast to the previous work, we introduce a new method for detecting the changes in  multivariate streaming data by using the geometric and clustering approach.

\section{Conclusions}\label{sec:Summary}
We have proposed a novel change detection method for streaming multivariate data by using clustering. Our change detection method uses the model fitting approach in which a change occurs when a new data item or block of data items  do not fit the existing clustering.
The salient features of clustering-based detector are that it can work well with the multivariate streaming data, and it is an automated change detection method for streaming data. Based on this change detector, we have presented a reactive algorithm for building and maintaining clustering emerging from the evolving data stream. The method for building and maintaining clustering in this paper can be extended for distributed environment.


\bibliographystyle{abbrv}
\bibliography{sigmod2012}

\end{document}